\magnification=1200
\baselineskip=20pt
\centerline{\bf A Hidden Connection between Lax Descriptions and
Superextensions}
\centerline{\bf of KdV Hierarchy}
\vskip 1.5cm
\centerline{Wen-Jui Huang}
\vskip 1cm
\centerline{Department of Electronic Engineering}
\centerline{National Lien Ho College of Technology and Commerce}
\centerline{Miao-Li, Taiwan, Republic of China}
\vskip 1.5cm
\centerline{\bf \underbar{Abstract}}
\vskip 1cm

A previously unnoticed connection between the Lax descriptions and the
superextensions of the KdV hierarchy is presented. It is shown that the
two different Lax descriptions of the KdV hierarchy come out naturally
from two different bihamiltonian superextensions of the KdV hierarchy.
Some implications of this observation are briefly mentioned.
\vfil\eject

\noindent{\bf 1. Introduction}

The KdV equation
$${\partial \over {\partial t}} u(x) = {1 \over 4} u'''(x) + {3 \over 2} u(x) u'(x) \eqno(1.1)$$
and its hierarchy had played an important role in the study of nonlinear
integrable partial differential equations[1-3]. A very economical way to express
the whole hierarchy is to use the Lax description:
$$ \eqalign{ {\partial \over {\partial t_n}} L &= [ (L^{n+{1 \over 2}})_+, L ]
   \cr
   L &= \partial^2 + u \cr} \eqno(1.2)$$

where $\partial = {\partial \over \partial x}$ and $ (\rm)_{\pm}$ refer to
terms which contain non-negative (negative) powers of $\partial$ in a
psuedodifferential operator. The KdV equation corresponds to $n=1$. Another
important property of the KdV hierarchy is that it has a bihamiltonian
structure[4]. In other words, the hierarchy can be described as a family of
hamiltonian flows in two distinct ways:
$$ {\partial \over {\partial t_n}} u(x) = \{ u(x), H_n \}_2 = \{ u(x), H_{n+1}
   \}_1 \eqno(1.3)$$
where
$$\eqalign{ \{ u(x), u(y) \}_2 &= [ {1 \over 2} \partial_x^3 + 2 u(x)
\partial_x  + u'(x) ] \delta(x-y) \cr
  \{ u(x), u(y) \}_1 &= 2 \partial_x \delta(x-y) \cr} \eqno(1.4)$$
In this approach the hamiltonians $H_n$'s are defined recursively from
the second equality of (1.3), which, in operator form, reads
$$ 2 \partial {{\delta H_{n+1}} \over {\delta u}} = [ {1 \over 2} \partial^3
    + 2 u \partial + u' ] {{\delta H_n} \over {\delta u}},\eqno(1.5)$$
and from the ``initial condition''
$$ H_0 = \int u dx \eqno(1.6)$$
From (1.5) a ``geometrical'' operator called recursion operator, $R$, can be
defined[5-7]
$$ R = 4 \left( {1 \over 2} \partial^3 + 2 u \partial + u' \right) \left(
   2 \partial \right)^{-1} = \partial^2 + 2 u + 2 \partial u \partial^{-1}
\eqno(1.7)$$
It can be shown[8] that $R$ can serve as a Lax operator in the sense that the
whole hierarchy can be put into the Lax form associated with $R$(with proper
rescales of $t_n$'s):
$${ \partial \over {\partial  t_n}} R = [ (R^{n+ {1 \over 2}})_+ , R ]
    \eqno(1.8)$$

For more than a decade the study of superextensions of various integrable
equations has been an active research area. The first superextension of
the KdV hierarchy is the Kupershmidt's sKdV hierarchy[9]. It is defined by
the supersymmetric extension of the bihamiltonian structure given by (1.3):
$$\eqalign{ \{ u(x), u(y) \}_2 &= [ {1 \over 2} \partial_x^3 + 2 u(x)
 \partial_x + u'(x) ] \delta (x-y) \cr
  \{ \phi(x), u(y) \}_2 &= [ {3 \over 2} \phi(x) \partial_x + \phi'(x) ]
     \delta (x-y) \cr
  \{ \phi(x), \phi(y) \}_2 &= [ \partial_x^2 + u(x) ] \delta (x-y) \cr
  \{ u(x), u(y) \}_1 &= 2 \partial_x \delta (x-y) \cr
  \{ \phi(x), u(y) \}_1 &= 0 \cr
  \{ \phi(x), \phi(y) \}_1 &= \delta (x-y) \cr} \eqno(1.9)$$

Anologuous to (1.3) the hamiltonians $H^{(s)}_n$'s in sKdV can be defined
recursively by
$$\eqalign{ \{u(x), H^{(s)}_{n+1} \}_1 &= \{ u(x), H^{(s)}_n \}_2 \cr
     \{ \phi(x), H^{(s)}_{n+1} \}_1 &= \{ \phi(x), H^{(s)}_n \}_2 \cr}
\eqno(1.10)$$
with the initial condition $H^{(s)}_0 = H_0$
It is interesting to note that the bracket $\{ \phi(x), \phi(y) \}_2$ is
precisely $L(x) \delta(x-y)$.  The appearance of the Lax operator $L$ given by
(1.2) in this bracket is not very surprising once the scale weight analysis
is considered. However, it does suggest a possible connection between the
hamiltonian structure of the sKdV hierarchy and the standard Lax description
(1.2) of the KdV hierarchy. Indeed, as we shall discuss later, the bihamltonian
description of the sKdV hierarchy leads to the Lax description (1.2) in a very
natural manner.

  What about the Lax description (1.8)? Is there a superextension of the KdV
hierarchy which gives this description in a similar way? The answer is yes!
We found that the subhierarchy (called the sKdV-B hierarchy) consisting of
the even flows of a recently discovered supersymmetric KdV
hierarchy[10] can be related to the Lax decription
(1.8) in a similar way.

We organize this paper as follows.  In Sec. 2 we discuss generally the
how a Lax description of an integrable equation can possibly
arise from a hamiltonian
superextension. In Sec. 3 we apply the idea in Sec. 2 to the KdV hierarchy
and show in details how the sKdV hierarchy leads to the Lax description (1.2).
In Sec. 4 we consider the same question for a superextension with grassmannian
odd hamiltonian structure. Then we consider the sKdV-B
hierarchy and show how the Lax description (1.8) arises in this case in Sec. 5.
Finally, we present our concluding remarks in Sec. 6.
\medskip

\noindent{\bf 2. Lax description and Superextension}

Now we consider an integrable equation which has a hamiltonian
description:
$$ {\partial \over {\partial t}} u(x) = \{ u(x), H \}	\eqno(2.1)$$
We assume further that this equation allows a hamiltonian superextension to
which a fermionic field $\phi(x)$ is added. The hamilton's equations of motion
are now
$$\eqalign{ {\partial \over {\partial t}} u(x) &= \{ u(x), \bar{H} \} \cr
  {\partial \over {\partial t}} \phi(x) &= \{ \phi(x), \bar{H} \} \cr}
   \eqno(2.2)$$
Here $\bar{H}$ is a superextension of $H$ in the sense that it reduces to $H$
when the fermionic field $\phi$ is set to zero. The question whether the
system (2.2) is integrable or not does not concern us here. As motivated by
the discussion following (1.10) we shall consider the $t$-evolution of
the bracket $\{ \phi(x), \phi(y) \}$:
$$ {\partial \over {\partial t}} \{ \phi(x), \phi(y) \} = \{ \{ \phi(x), \phi(y)
\}, \bar{H} \} \eqno(2.3)$$
Using the super Jacobi identity[11]:
$$ (-1)^{|A||C|} \{ \{ A, B \}, C \} + (-1)^{|B||A|} \{ \{ B, C \}, A \}
   + (-1)^{|C||B|} \{ \{ C, A \}, B \} = 0 \eqno(2.4)$$
where $|A|=0(1)$ if $A$ is bosonic (fermionic), we can write (2.3) as
$$ \eqalign{ {\partial \over {\partial t}} \{ \phi(x), \phi(y) \} &=
    \{ \phi(x), \{ \phi(y), \bar{H} \} \} + \{ \{ \phi(x), \bar{H} \}, \phi(y)
\} \cr
&= \int dz \{ \phi(x), \phi(z) \} {\delta \over {\delta
\phi(z) }} \{ \phi(y), \bar{H} \} \cr
  &\quad + \int dz \{ \phi(z), \phi(y) \} {\delta \over
{\delta \phi(z) }} \{ \phi(x), \bar{H} \}  \cr
  &\quad + S \cr} \eqno(2.5)$$
({\it Note:} All functional derivatives in papers are left-derivatives)
Here $S$ is the collection of terms which will vanish when the condition
``$\phi=0$'' is imposed.
We write
$$ \eqalign{ \{ \phi(x), \phi(y) \}|_{\phi=0} &\equiv L(x) \delta(x-y) \cr
    {\delta \over {\delta \phi(y)}} \{ \phi(x), \bar{H} \}|_{\phi=0}
     &\equiv M(x) \delta(x-y) \cr} \eqno(2.6)$$
Then
$$ {\delta \over { \delta \phi(z)}} \{ \phi(y),
   \bar{H} \}|_{\phi=0} =
   M(y) \delta(y-z) = M^*(z) \delta (z-y)  \eqno(2.7)$$
where $M^*$ denotes the adjoint of $M$ ({\it Recall:} $\partial^* = -\partial$
and $f^* = f$. ) Now taking the bosonic limit of (2.5) (i.e., imposing $\phi=0$
) and using (2.6)-(2.7) we get
$$ {\partial \over {\partial t}} L = M L + L M^*  \eqno(2.8)$$
If it happens that $M$ is anti-self adjoint, that is,
$$ M^* = -M \eqno(2.9)$$
then (2.8) becomes the Lax equation:
$$ {\partial \over {\partial t}} L = [ M, L ] \eqno(2.10)$$
The remaining question is then whether or not an $\bar{H}$ can be found so
that the operator $M$ defined by (2.5) is anti-self adjoint. The answer to
this question, of course, depends on the nature of the extended system (2.2).
We shall demonstrate in the following section that it can be done for
the superextension of the KdV hierarchy based on the super hamiltonian
structure $\{ ,  \}_2$ given by (1.9).

As an ending remark we like to point out that the hamiltonian $\bar{H}$ which
makes (2.9) satified is certainly not unique. This is because the limit
$\phi=0$ is taken in the definition of $M$. Hence, those terms in $\bar{H}$,
which are at least quartic in $\phi$, would not enter the expression for $M$.
\medskip

\noindent{\bf 3. Lax description of the KdV hierarchy from the sKdV hierarchy}

In this section we shall apply the idea in section 2 to the KdV hierarchy.
We like to extend the hierarchy by adding a fermionic field $\phi$.
Since we are interested in hamiltonian superextensions defined by the bracket
$\{, \}_2$ given by (1.9), we only have to consider the superextensions of
the hamiltonians $H_n$'s defined recursively by (1.5).
Let us begin with the zeroth flow of (1.2). The hamiltonian for this flow
is simply $H_0$ given by (1.6). Quite obviously there is no nontrivial
superextension of $H_0$. In other words, we have to take
$$ \bar{H}_0 = H_0 = \int u dx \eqno(3.1)$$
Putting $\bar{H}_0$ into (2.6) and using (1.9) we get
$$ M_0 = \partial \eqno(3.2)$$
which is evidently anti-self adjoint. Moreover,
$$ M_0 = (L^{1 \over 2})_+ \eqno(3.3)$$
where $L= \partial^2 + u$ as in section 1.
Combining (2.10) with (3.3) then gives
$$ {\partial \over {\partial t_0}} L = [ M_0, L ] = [ (L^{1 \over2})_+, L ]
   \eqno(3.4)$$
The equation (3.4) is precisely (1.2) with $n=0$.
Next we consider the hamiltonian for the first flow (KdV equation)
$$ H_1 = {1 \over 4} \int u^2 dx \eqno(3.5)$$
This hamiltonian allows an one-parameter family of superextensions
$$ \bar{H}_1 = {1 \over 4} \int ( u^2 + a \phi \phi' ) dx \eqno(3.6)$$
The corresponding operator defined by (2.6) is computed to be
$$ M_1 =  {1 \over 2} a \partial^3 + {1 \over 2} (1+a) u \partial +
	  {3 \over 4} u'   \eqno(3.7)$$
One can show easily that $M_1$ is anti-self adjoint if and only if $a=2$.
With this value of $a$ one can check
$$\eqalign{ M_1 &= \partial^3 + {3 \over 2} u \partial + {3 \over 4} u' \cr
		&= ( L^{3 \over 2} )_+ \cr} \eqno(3.8)$$
We thus have recovered (1.2) with $n=1$. As the final explicit example we
consider the second flow. The corresponding hamiltonian allows a two-parameter
family of superextensions:
$$ \bar{H}_2 = {1 \over 8} \int [ u^3 + {1 \over 2} u u'' + a u \phi \phi'
	      + b \phi \phi''' ] dx \eqno(3.9)$$
In this case, the second equation of (2.6) gives
$$\eqalign{ M_2 &= {b \over 4} \partial^5 + {1 \over 4} (a+b) u \partial^3
 + {5 \over 8} a u' \partial^2 + [ {1 \over 8} (2a+3) u^2 + {1 \over 8} (4a+1)
 u''] \partial \cr
&\quad + {1 \over 16} (2a+3) u''' + {1 \over 8} (a+9) u u' \cr} \eqno(3.10)$$
Requiring $M_2$ to be anti-self adjoint gives the unique solution: $a=6$ and
$b=4$. Thus
$$\eqalign{ M_2 &= \partial^5 + {5 \over 2} u \partial^3 + {15 \over 2} u'
  \partial^2 + ({15 \over 8} u^2 + {25 \over 8} u'') \partial + {15 \over 16}
u''' + {15 \over 8} u u' \cr
   &= (L^{5 \over 2} )_+ \cr} \eqno(3.11)$$
as expected.

The above explicit calculations should have convinced one to expect that the
Lax description (1.2) for the whole KdV hierarchy can be reproduced in this
manner. In other words, we expect that for each $n \geq 0$ a suitable
hamiltonian $\bar{H}_n$ can be found so that the resulting $M_n$ is anti-self
adjoint and that
$$ M_n = ( L^{n + { 1 \over 2}} )_+  \eqno(3.12)$$

We now are going to show that the above expectation and, especially, (3.12)
are indeed true. Our proof comes from the observation that for $n=0,1,2$
the equality
$$ \bar{H}_n = H^{(s)}_n \eqno(3.13)$$
actually holds. Here $H^{(s)}_n$'s are the hamiltonians for the
sKdV hierarchy[9],
which have been defined recursively by (1.10). As a matter of fact, taking
(3.13) for all values of $n$ does provide an ansatz to our problem. To see
this, we use the second of (1.10) to get
$$\eqalign{ M_n (x) \delta(x-y) &\equiv
{\delta \over {\delta \phi(y)}} \{ \phi(x), H^{(s)}_n \}_2|_{\phi=0} \cr
&= {\delta \over {\delta \phi(y)}} \{ \phi(x), H^{(s)}_{n+1} \}_1|_{\phi=0}
\cr
&= {{\delta^2 H^{(s)}_{n+1}} \over {\delta \phi(y) \delta \phi(x)}}|_{\phi=0}
\cr} \eqno(3.14)$$
Since
$$ {\delta^2 \over { \delta \phi(y) \delta \phi(x)}} = -
   {\delta^2 \over { \delta \phi(x) \delta \phi(y)}}
   \eqno(3.15)$$
it follows immediately from (3.14) that
$$ M_n(x) \delta(x-y) = - M_n(y) \delta(x-y) \eqno(3.16) $$
or, equivalently,
$$ M_n^* = - M_n \eqno(3.17)$$
We therefore have shown that for all $n \geq 0$
$$ {\partial \over {\partial t_n}} L = [ M_n, L ] \eqno(3.18)$$
and that $M_n$ is given by (3.14). It remains to prove (3.12). To this end,
we start with the explicit form of the second equality of (1.10):
$${{\delta H^{(s)}_{n+1}} \over {\delta \phi(x)}} =
\left( \partial_x^2 + u(x) \right) {{\delta H^{(s)}_n} \over
{\delta \phi(x)}} + \left( {3 \over 2} \phi(x) \partial_x
 + \phi'(x) \right) {{\delta H^{(s)}_n} \over {\delta u(x)}}  \eqno(3.19)$$
Differentiating both sides of (3.19) with respect to $\phi(y)$ and setting
$\phi$ to zero yields
$${{\delta^2 H^{(s)}_{n+1}} \over {\delta \phi(y) \delta \phi(x)}}
|_{\phi=0} = \left( \partial_x^2 + u(x) \right)
{{\delta^2 H^{(s)}_n} \over {\delta \phi(y) \delta \phi(x) }}|_{\phi=0}
 + \left( {3 \over 2} \delta(x-y) \partial_x + \delta'(x-y) \right)
 {{\delta H_n} \over {\delta u(x)}}  \eqno(3.20)$$
Combining (3.14), (3.20) and the following relation
$$ 2 \partial_x {{\delta H^{(s)}_{n+1}} \over {\delta u(x)}}  = \{ u(x), H_{n+1} \}_1
   = {\partial \over {\partial t_n}} L(x) = [ M_n(x), L(x) ] \eqno(3.21)$$
we obtain a recursion relation for $M_n$'s:
$$ M_n = L M_{n-1} + {1 \over 2} \left( \partial^{-1} [ M_{n-1}, L ] \right)
  \partial + {3 \over 4} [ M_{n-1}, L ] \eqno(3.22)$$
The validity of (3.11) then follows from the facts that $M_0= \partial = (L^
{1 \over 2})_+$ and that $(L^{n + {1 \over 2}})_+$'s satisfy a recursion
relation identical to (3.22) (See Appendix A for a proof).

We have shown that the Lax description (1.2) of the KdV hierarchy can be
reproduced from the Kupershmidt's sKdV hierarchy in a pretty natural way.
This analysis also provides an insight into the structure of the hamiltonians,
$H^{(s)}_n$'s. From (3.12) and (3.14) we deduce that
$$ H^{(s)}_n = H_n + {1 \over 2} \int dx \left[ \phi (L^{n - {1 \over 2}})_+
\phi + O(\phi^4) \right] \eqno(3.23)$$
Here $O(\phi^4)$ represents the collection of all terms which are at least
quartic in $\phi$. As mentioned in the end of last section, the terms in
$O(\phi^4)$ play no role in the definition of $M_n$. Hence we can simply take
$\bar{H}_n$ to be the sum of the first two terms on the right hand side of
(3.23).
\medskip
\vfil\eject

\noindent{\bf 4. Odd hamiltonian structure}

It was recently discovered that there exists a new supersymmetric extension of
the KdV hierarchy. This hierarchy (called sKdV-B hierarchy) is essential the
KdV hierarchy, where the
KdV field is replaced by an even superfield. One interesting feature of this
hierarchy is that it is based on an odd hamiltonian structure, instead of
an even
one. We shall show that applying the previous idea to this hierarchy enables
us to reproduce another Lax description of the KdV hierarchy, namely, (1.8).
In this section we shall generalize the idea used in section 2 to an
odd hamiltonian structure.

Let us consider a bosonic evolution equation
$$ {\partial \over {\partial t}} u(x) = F[u(x)] \eqno(4.1)$$
Assume that this equation can be extended by a fermionic field $\psi$ in such a
way that the extended system is hamiltonian with respect to an odd hamiltonian
structure:
$$\eqalign{ {\partial \over {\partial t}} u(x) &= ( u(x), K ) \cr
	    {\partial \over {\partial t}} \psi(x) &= ( \psi(x), K ) \cr}
 \eqno(4.2)$$
Obviously the hamiltonian $K$ must be odd in order that the first equality
of (4.2) can
give a nontrivial equation in the $\psi = 0$ limit. Before proceeding further
we list first a few important properties of an odd hamiltonian structure for
later uses:
$$\eqalign{ | ( F, G ) | &= |F| + |G| + 1 \cr
	     ( F, G ) &= - (-1)^{(|F|+1)(|G|+1)} ( G, F ) \cr
	     (-1)^{(|F|+1)(|H|+1)} ( F, ( G, H )) &+ \quad {\it cyclic
\quad
permutations} =0 \cr} \eqno(4.3)$$
Since both of the brackets $(u,u)$ and $(\psi, \psi)$ are odd, each of them
will have a trivial $t$-evolution in the $\psi = 0$ limit. We consider
the $t$-evolution of $(u, \psi)$:
$$\eqalign{ {\partial \over {\partial t}} ( u(x), \psi(y) ) &=
((u(x), \psi(y)), K ) \cr
&= ( u(x), (\psi(y), K )) + ( ( u(x), K ), \psi(y) ) \cr} \eqno(4.4)$$
Taking $\psi=0$ then gives
$$ \eqalign{ {\partial \over {\partial t}} (u(x), \psi(y))|_{\psi=0} &=
   \int dz \left[ ( u(x), \psi(z) ) {\delta \over {\delta \psi(z)}} ( \psi(y),
 K ) \right]_{\psi=0} \cr
    &\qquad + \int dz \left[ ( u(z), \psi(y)) {\delta \over {\delta u(z)}}
    ( u(x), K ) \right]_{\psi=0} \cr} \eqno(4.5)$$
Writing
$$\eqalign{ ( u(x), \psi(y) ) |_{\psi=0} &\equiv R(x) \delta(x-y) \cr
   {\delta \over {\delta \psi(x) }} ( \psi(y), K ) |_{\psi = 0} &\equiv N(x)
\delta(x-y) \cr
{\delta \over {\delta u(y)}} (u(x), K)|_{\psi=0} &\equiv M(x) \delta(x-y) \cr}
\eqno(4.6)$$
we have
$$ {\partial \over {\partial t}} R =  R N + M R \eqno(4.7)$$
When
$$ N = - M \eqno(4.8)$$
or, equivalently,
$${\delta \over {\delta \psi(x)}} ( \psi(y), K )|_{\psi=0} = -
  {\delta \over {\delta u(y)}} ( u(x), K )|_{\psi=0}  \eqno(4.9)$$
the operator equation (4.7) become of the Lax form:
$$ {\partial \over {\partial t}} R = [ M, R ] \eqno(4.10)$$
Hence we have a Lax description for the bosonic equation (4.1). It  is worth
noting that (4.1), (4.2) and the last of (4.6) together imply
$$ M = {\delta \over {\delta u}} F[u] \eqno(4.11)$$
that is, $M$ is simply the Frechet derivative of $F[u]$.

In the following section we shall check (4.9) explicitly for
the sKdV-B hierarchy and show that (4.10) and (4.11) precisely give the
Lax description (1.8) of the KdV hierarchy.
\medskip
\vfil\eject

\noindent{\bf 5. A Lax description from the sKdV-B hierarchy}

The hamiltonians, $K_n$'s, of the sKdV-B hierarchy can be defined recursively
by
$$\eqalign{ ( u(x), K_{n+1} )_1 &= ( u(x), K_n )_2  \cr
	( \psi(x), K_{n+1} )_1 &= ( \psi(x), K_n )_2  \quad (n \geq 1) \cr}
\eqno(5.1)$$
together
with the initial condition
$$ K_1 = {1 \over 4} \int dx u' \psi \eqno(5.2)$$
Here $( , )_{1,2}$ are two odd hamiltonian structures defined as
$$ \eqalign{ ( \psi(x), \psi(y) )_2 &= 2 ( \partial_x^{-1} \psi'(x) + \psi'(x)
    \partial_x^{-1} ) \delta (x-y) \cr
  ( u(x), \psi(y) )_2 &= \left( \partial_x^2 + 2 u(x) + 2 \partial_x u(x)
\partial_x^{-1} \right) \delta (x-y) \cr
  ( u(x), u(y) )_2 &= 0 \cr
  ( u(x), \psi(y) )_1 &= 4 \delta(x-y) \cr
  ( u(x), u(y) )_1 &= ( \psi(x), \psi(y) )_1 = 0 \cr} \eqno(5.3)$$
One should note that the recursion relations (5.1) start from $n=1$. The zeroth
flow of the KdV hierarchy:
$$ {\partial \over {\partial t}} u(x) = {\partial \over {\partial x}} u(x)
    \eqno(5.4)$$
is not included in this hierarchy since it is never a $\psi=0$ limit of a
hamiltonian equation of the form
$$ {\partial \over {\partial t}} u(x) = ( u(x), K )_2 \eqno(5.5)$$
Note also that the normalization of $K_1$ has been chosen so that the
hamiltonian equation (5.5), with
$K$ replaced by $K_1$, gives precisely the KdV equation (1.1).

Our first task is to check if $K_n$'s satisfy (4.9). Using (5.1) and (5.3)
we arrive at
$$ \eqalign{ {\delta \over {\delta \psi(x)}} ( \psi(y) , K_n )_2 |_{\psi=0}
   &= {\delta \over {\delta \psi(x)}} ( \psi(y), K_{n+1} )_1 |_{\psi=0} \cr
   &= -4 {{\delta^2 K_{n+1}} \over {\delta \psi(x) \delta u(y)}}|_{\psi=0} \cr
   {\delta \over {\delta u(y)}} ( u(x), K_n )_2 |_{\psi=0} &=
    {\delta \over {\delta u(y)}} ( u(x), K_{n+1} )_1 |_{\psi=0} \cr
   &= 4 {{\delta^2 K_{n+1}} \over {\delta u(y) \delta \psi(x)}}|_{\psi=0} \cr}
 \eqno(5.6)$$
which obviously verifies (4.9).
Now if we write the KdV hierarchy as
$$ {\partial \over {\partial t_n}} u(x) = F_n [u(x)] \eqno(5.7)$$
then we conclude from (4.10) and (4.11) that the KdV hierarch has the following
Lax description
$$ {\partial \over {\partial t_n}} R = [ M_n, R ] \eqno(5.8)$$
with
$$ \eqalign{ R = \partial^2 + 2 u + 2 \partial u \partial^{-1} \cr
	     M_n \equiv {\delta \over {\delta u}} F_n [u] \cr} \eqno(5.9)$$
This description is a well known result in the geometrical approach to the
KdV hierarchy. The operator $R$ is known as the recursion operator. It can be
shown (see Appendix B ) that
$$ M_n = 4^{-n} ( R^{n+ {1 \over 2}})_+ \eqno(5.10)$$
In fact, applying (5.9) to the first and the second flows of (1.2) we readily
verify
$$\eqalign{ M_1 &= {1 \over 4} ( \partial^3 + 6 u \partial + 6 u' ) \cr
		&= {1 \over 4} ( R^{3 \over 2} )_+ \cr} \eqno(5.11)$$
and
$$\eqalign{ M_2 &= {1 \over 16} \left( \partial^5 + 10 u \partial^3 + 20 u'
\partial^2 + ( 20 u'' + 30 u^2 ) \partial + 10 u''' + 60 u u' \right) \cr
       &= {1 \over 16} ( R^{5 \over 2} )_+ \cr} \eqno(5.12)$$
as expected.

The relation (5.10) suggests us to rescale the evolution parameters $t_n$'s
as follows
$$ t_n \longrightarrow 4^n t_n \eqno(5.13)$$
The the Lax equation (5.8) then becomes (1.8):
$$ {\partial \over {\partial t_n}} R = [ 4^n M_n, R ] = [ (R^{n+ {1 \over
2}})_+, R ] \eqno(5.14)$$
As claimed we have shown that sKdV-B hierarchy leads naturally to the Lax
description of the KdV hierarchy, which is in terms of the recursion operator
$R$.
\medskip

\noindent{\bf 6. Concluding Remarks}

We have demonstrated how the Kupershimidt's sKdV hierarchy leads naturally
to the Lax description (1.2) for the KdV hierarchy. In our derivation there
are two key ingredients. One is the super Jacobi identity which has been used
to obtain (2.8), a preliminary form of the Lax description. The other is the
the recursive definition (1.10) of the hamiltonians in the sKdV hierarchy.
The relation (1.10) together with the form of $\{ \phi(x), \phi(y) \}_1$
guarantees anti-self adjointness of the corresponding $M_n$'s and hence
promotes (2.8) to a genuine Lax description.

We have also discussed a similar connection between the alternate Lax
description (1.8), which arises from the geometrical approach to the KdV
hierarchy, and the recently discovered sKdV-B hierarchy appearing in a theory
of 2-d quantum gravity. An interesting feature of this hierarchy is that it is
based on an odd hamiltonian structure. However, it's perhaps fair to say that
the connection in this case is quite expectable since the sKdV-B hierarchy is
obtained from the geometrical approach to the KdV hierarchy by a simple
replacement of the original bosonic field by a even superfield. Nevertheless,
as seen in Sec. 4 and Sec. 5, this example again shows the importance of the
super Jacobi identity and the recursion relation for the hamiltonians in the
derivation of Lax description.

It is easy to generalize the discussion in Sec. 2 to the systems with arbitrary
number of fields. In this more general situation the operators $L$ and $M$
appearing in (2.10) become differential operators with matrix-valued
coefficients. As a result of the antisupersymmetric property of a super-
hamiltonian structure, $L$ must be self adjoint. Hence, it seems that our
Lax description (2.10) is quite restrictive. It is not clear whether or not
some other nontrivial examples exist. If so, then
the connection discussed in this paper does suggest a possibe method to
construct
 Lax descriptions for such systems: one tries to find a
superextension and then applies the method of Sec. 2. Of course, finding a
superextension is not an easy task. The method may even be more difficult than
other conventional approaches. However, we do suspect that it could be useful
in some other systems.

The remarks in the previous paragraph apply equally well to the discussion in
Sec. 4 except that now the operator $R$ has no definite symmetry property.
\medskip

\noindent{\bf Appendix A: A recursion reltaion for $( L^{n + {1 \over 2}}
)_+$'s}

In this appendix we like to show that $( L^{ n + {1 \over 2}} )_+$'s satisfy
the recursion relation (3.22) as $M_n$'s do.

We first write
$$ L^{n - {1 \over 2}} = ( L^{n - {1 \over 2}} )_+ + \partial^{-1} a_1 +
   \partial^{-2} a_2 + \dots \eqno(A.1)$$
Then we compute to get
$$ ( L^{ n + {1 \over 2}} )_+ = ( L  L^{n - {1 \over 2}} )_+ =
   L ( L^{n - {1 \over2}} )_+ + a_1 \partial
       + a'_1 + a_2 \eqno(A.2)$$
and
$$ ( L^{ n + {1 \over 2}} )_+ = ( L^{n - {1 \over 2}}  L )_+ =
   ( L^{n - {1 \over 2}} )_+ L + a_1 \partial - a'_1 + a_2  \eqno(A.3)$$
Equating (A.2) with (A.3) yields
$$ 2 a'_1 = [ ( L^{ n - {1 \over 2}} )_+, L ] \eqno(A.4)$$
We can use (A.4) to derive a relation between $a_1$ and $a_2$ by comparing
both sides of (A.4):
$$\eqalign{ 2 a'_1 &= [ ( L^{n - {1 \over 2}} )_+, L ] =
		      [ L, ( L^{n - {1 \over 2}} )_- ] \cr
       &= [ \partial + u, \partial^{-1} a_1 + \partial^{-2} a_2 + \dots ]
\cr
  &= 2 a'_1 + ( 2 a'_2 - a''_1 ) \partial^{-1} + \dots \cr} \eqno(A.5)$$
We have
$$ a_2 = {1 \over 2} a'_1 \eqno(A.6)$$
Substituting (A.4) and (A.6) into the right hand side of (A.2) we obtain the
desired recursion relation:
$$ ( L^{n + {1 \over 2}} )_+ = L ( L^{n - {1 \over 2}} )_+ {1 \over 2}
  \left( \partial^{-1} [ ( L^{ n - {1 \over 2}} )_+, L ] \right) \partial
   + {3 \over 4} [ ( L^{n - {1 \over 2}} )_+ , L ] \eqno(A.7)$$

\medskip

\noindent{\bf Appendix B: A proof of (5.10) }

Performing functional derivative with respect to $u(y)$ on both sides of the
first relation in (5.1) we have
$$\eqalign{ 4 {{\delta^2 K_{n+1}} \over {\delta u(y) \delta \psi(x)}} &= \left(
\partial_x^2 + 2 u(x) + 2 \partial_x u(x) \partial^{-1} \right)
 {{\delta^2 K_n} \over {\delta u(y) \delta \psi(x)}} + 2 \left[ \partial_x^{-1}
 {{\delta K_n} \over {\delta \psi(x)}} \right] \partial_x \delta(x-y) \cr
  &\qquad + 4 {{\delta K_n} \over {\delta \psi(x)}} \delta(x-y) \cr}
 \eqno(B.1)$$
Note that (before taking the rescale operation (5.13))
$$ {\partial \over {\partial t_{n-1}}} u(x) = ( u(x), K_n )_1 =
     4 {{\delta K_n} \over {\delta \psi(x)}} \eqno(B.2)$$
and
$$ {\partial \over {\partial t_{n-1}}} R = [ M_{n-1} ,R] \eqno(B.3)$$
together imply
$$ {{\delta K_n} \over {\delta \psi(x)}}|_{\psi=0} = {1 \over 4} {\partial
\over	 {\partial t_{n-1}}} u(x) = {1 \over 16} [ M_{n-1}, R ]_0 \eqno(B.4)$$
where $(A)_0$ denotes the zeroth order term of a pseudodifferential operator
$A$. On the other hand, from the definition of $M_n$ (see (4.6)) and from (5.1)
we have
$$ M_n (x) \delta (x-y) = 4 {{\delta^2 K_{n+1}}\over {\delta u(y) \delta
\psi(x)}}|_{\psi=0} \eqno(B.5)$$
Putting (B.4) and (B.5) into the $\psi=0$ limit of (B.1) gives a recursion
relation for $M_n$'s:
$$ M_n = {1 \over 4} R M_{n-1} + {1 \over 8} \left( \partial^{-1} [ M_{n-1}, R
]_0  \right) \partial + {1 \over 4} [ M_{n-1}, R ]_0 \eqno(B.6)$$

Next we need to show that (B.6) is also a recursion relation for
$(R^{n+ {1 \over 2}})_+$'s.
Writing
$$ R^{n - {1 \over 2}} = ( R^{n - {1 \over 2}} )_+ + \partial^{-1} b_1 +
   \partial^{-2} b_2 + \dots \eqno(B.7)$$
we find
$$ ( R^{n + {1 \over 2}} )_0 = ( R R^{n - {1 \over 2}} )_0 = ( R (R^{n - {1
\over 2}} )_+ )_0 + b'_1 + b_2 \eqno(B.8)$$
and
$$ ( R^{n + {1 \over 2}} )_0 = ( R^{n -{1 \over 2}} R )_0 = ( (R^{n - {1 \over
2}} )_+ R )_0  - b'_1 + b_2  \eqno(B.9)$$
Eqating (B.8) with (B.9) gives
$$ 2 b'_1 = [ (R^{n - {1 \over 2}} )_+, R ]_0 \eqno(B.10)$$
Since $b_1 = Res(R^{n - {1 \over 2}} )$, we thus have
$$ Res(R^{n - {1 \over 2}}) = {1 \over 2} \left( \partial^{-1} [ (R^{n - {1
\over 2}})_+, R] \right) \eqno(B.11)$$
In ref.[8] it has been shown that  $R$ satisfies
$$ (R^{n + {1 \over 2}} )_+ = R (R^{n - {1 \over 2}})_+ + \left( Res(R^{n - {1
\over 2}}) \right) \partial + 2 \left( \partial Res(R^{n - {1 \over 2}})
\right) \eqno(B.12)$$
Combining (B.11) with (B.12) finally yields
$$ (R^{n + {1 \over 2}} )_+ = R (R^{n - {1 \over 2}} )_+ + {1 \over 2}
   \left( \partial^{-1} [ (R^{n - {1 \over 2}})_+, R ]_0 \right) \partial
   + [ (R^{n - {1 \over 2}} )_+, R ]_0 \eqno(B.13)$$
Comparing (B.6) to (B.13) we conclude immediately that $4^n M_n$'s and $R^{n +
{1 \over 2}}$'s obey the same recursion relation.  This result together with
(5.11) completes the proof of (5.10).

\vfil\eject

\noindent{\bf References:}
\item{[1]} L.D. Faddeev and L.A. Takhtajan, {\it Hamiltonian Methods in
the Theory of Solitons} (Springer-Verlag, Berlin, 1987).
\item{[2]} A. Das, {\it Integrable Models} (World Scientific, Singapore,
1988).
\item{[3]} L. Dickey, {\it Soliton Equations and Hamiltonian Systems } (World
Scientific, Singapore, 1991).
\item{[4]} F. Magri, J. Math. Phys. {\bf 19}, 1156 (1978).
\item{[5]} H.H. Chen, Y.C. Lee and C.S. Liu, Phys. Script {\bf 20},
 490 (1979).
\item{[6]} F. Magri, in {\it Nonlinear Evolution Equations and
Dynamical Systems}, eds. M. Boiti, F. Pempinelli and G. Soliani, {\it
Lecture Notes in Physics, Vol. 120} (Springer, 1980).
\item{[7]} P.J. Olver, {\it Applications  of Lie Groups to Differential
Equatiions, Graduate Texts in Mathematics, Vol. 107} (Springer, 1986).
\item{[8]} J.C. Brunelli and A. Das, Mod. Phys. Lett. {\bf A10}, 931 (1995).
\item{[9]} B.A. Kupershmidt, Phys. Lett. {\bf A102}, 213 (1984).
\item{[10]} J.M. Figueroa-O'Farrill and S. Stanciu, Phys. Lett. {\bf
B316}, 282 (1993).
\item{[11]} B. DeWitt, {\it Supermanifolds} (Cambridge University Press,
New York, 1984).

\end